\documentclass{JHEP3}
\usepackage{graphicx}
\usepackage{amsmath}
\usepackage{amsfonts}
\usepackage{amssymb}
\usepackage{epsfig}
\newcommand{\be}{\begin{equation}}
\newcommand{\ee}{\end{equation}}
\newcommand{\ben}{\begin{eqnarray}}
\newcommand{\een}{\end{eqnarray}}
\newcommand{\bb}{\bibitem}

\newcommand{\wt}{\widetilde}
\title{Brane Structure from a Scalar Field in Warped Spacetime}
\author{D. Bazeia, C. Furtado\\
Departamento de F\'\i sica, Universidade Federal da Para\'\i ba,
Caixa Postal 5008, 58051-970, Jo\~ao Pessoa Para\'\i ba, Brazil\\
e-mail: bazeia@fisica.ufpb.br, furtado@fisica.ufpb.br}
\author{A.R. Gomes\\
Departamento de F\'\i sica, Universidade Federal da Para\'\i ba,
Caixa Postal 5008, 58051-970, Jo\~ao Pessoa Para\'\i ba, Brazil\\
Departamento de Ci\^encias Exatas, Centro Federal de
Educa\c c\~ao Tecnol\'ogica do Maranh\~ao, 65025-001 S\~ao Lu\'\i s
Maranh\~ao, Brazil\\
e-mail: argomes@fisica.ufpb.br}
\abstract{We deal with scalar field coupled to gravity in five dimensions
in warped geometry. We investigate models described by potentials
that drive the system to support thick brane solutions that engender
internal structure. We find analytical expressions for the brane solutions,
and we show that they are all linearly stable.\\

\vspace{1cm}

}
\begin{document}
\maketitle

%%%%%%%%%%%%%%%%%%%%%%%%%%%%%%%%%%
Branes in higher dimensional theories provide an interesting new procedure
for resolving problems involving cosmological constant and hierarchy
\cite{add,rs}. The brane scenario that we are interested in the present letter
is driven by a single real scalar field. It engenders the interesting feature
of localizing massless graviton in the brane, efficiently reproducing four
dimensional gravity in the brane. The study of scalar fields coupled to
gravity in warped geometries has gained renewed attention
recently \cite{gw,gt,grs,df,g,ce}. In particular, in Ref.~{\cite{c}} one
investigates the splitting of thick branes due to a first-order phase
transition in a warped geometry. The work of Ref.~{\cite{c}} considers
a complex scalar field coupled to gravity, and it shows that when the
temperature approaches its critical value, an interface that interpolates two
bulk phases breaks into two separated interfaces, giving rise to a new phase
which appears in between these two interfaces, leading to an effect which
is known in condensed matter as complete wetting --- see also Ref.~{\cite{chw}}
for other details in the subject.

In the present work we show that similar results also appear at zero
temperature, in a simpler model, which depends on a single real scalar field,
coupled to gravity in five dimensions in a warped geometry. The results
seem to be of compelling interest to high energy physics, and are depicted
in the figures below. The potential that controls the scalar
field depends on odd integers in a very specific manner \cite{bmm},
allowing the appearance of thick branes that host internal structure,
in the form of a layer of a new phase squeezed in between two separate
interfaces. The potential that we investigate here was first studied in
Ref.~{\cite{bmm}}, and there it was shown that although it depends on a
single real scalar field, it supports 2-kink solutions of the BPS type,
which are stable solutions that solve first order differential equations
and engender the feature of being composed of two standard kinks.
These topological defects are richer than the standard topological
kinks, and they may be seen as defects that host internal structures. This
feature establishes the main motivation of this work, in which we explore the
possibility to offer a brane scenario where the brane entraps internal
structure. The idea is similar to the investigations done in
\cite{{m,brs}} in flat spacetime, in models described by two
real scalar fields, and in \cite{gp} in curved spacetimes, in
brane scenarios involving two higher dimensions. In the present
work, however, we shall deal with very specific models, described
by a single real scalar field, which couples with gravity
in warped spacetime in one higher dimension.

To show how the above potential behaves in the brane scenario,
we follow Refs.~{\cite{df,ce}} and we consider the action
\be
S=\int d^4x\,dy\sqrt{|g|}\Bigl[-\frac14 R+
\frac12\partial_a\phi\partial^a\phi-V(\phi)\Bigr]
\ee
where $g=\det(g_{ab})$ and the metric is
\be
ds^2=g_{ab}dx^adx^b=e^{2A}\eta_{\mu\nu}dx^{\mu}dx^{\nu}-dy^2
\ee
where $a,b=0,1,2,3,4,$ $\mu,\nu=0,1,2,3,$ $\;\eta_{\mu\nu}=(1,-1,-1,-1)$
and $e^{2A}$ is the warp factor. We suppose that the scalar field and the
warp factor only depend on the extra coordinate $y$. In this case the
equations of motion are
\ben
\phi^{\prime\prime}+4A^\prime\phi^\prime&=&
\frac{dV}{d\phi}
\\
A^{\prime\prime}&=&-\frac23\,\phi^{\prime2}
\\
A^{\prime2}&=&\frac16\phi^{\prime2}-\frac13 V(\phi)
\een
where prime stands for derivative with respect to $y$. Our model
is given by
\be\label{gpot}
V_p(\phi)=\frac18 \left(\frac{dW_p}{d\phi}\right)^2-\frac13 W^2_p
\ee
where \cite{bmm}:
\be\label{w}
W_p(\phi)=\frac{2p}{2p-1}\,\phi^{\frac{2p-1}{p}}-\frac{2p}{2p+1}\,
\phi^{\frac{2p+1}{p}}
\ee
The parameter $p$ is odd integer and controls the way the
scalar field self-interacts. For clarity, in Fig.~[1] we depict the potential
$V_p(\phi)$ for $p=1,3,5.$ 

The function $W_p(\phi)$ was first introduced in
Ref.~{\cite{bmm}} in flat spacetime. It was inspired in the work on deformed
defects \cite{blm}, and it appears with the help of the
deformation prescription there proposed. To see how this is done,
we suppose that the spacetime is flat. In this case
the potential is given by ${\wt V}(\phi)=(1/2)(d{\wt W}/d\phi)^2$. We consider
the function ${\wt W}(\phi)=\phi-\phi^3/3,$ which identifies the standard
$\phi^4$ model. We now follow \cite{blm} and introduce the
function $f(\phi)=\phi^{1/p}$, which is well-defined for every $\phi,$
for $p$ odd integer. This function can be used to change
${\wt W}(\phi)=\phi-\phi^3/3$ to a new model, defined by ${\wt W}_p(\phi)$.
The deformation prescription leads to the general result
$(df/d\phi)(d{\wt W}_p/d\phi)=(d{\wt W}/d\phi)[\phi\to f(\phi)],$
where in the right hand side one needs to change $\phi$ by $f(\phi),$ after
taking the derivative of ${\wt W}(\phi).$ For the case under consideration
we get ${\wt W}_p(\phi)=(p/2)W_p(\phi),$ that is, $(p/2)$ times
the expression in Eq.~(\ref{w}), which is used to define the potential
$V_p(\phi).$

%%%%%%%%%%%%%%%%%%%%%%%%%%%%%%%%%%%%%%%%%%%%%%%%%%%%%%%%%%%%%%%%%%%%%%%
\EPSFIGURE[ht!]{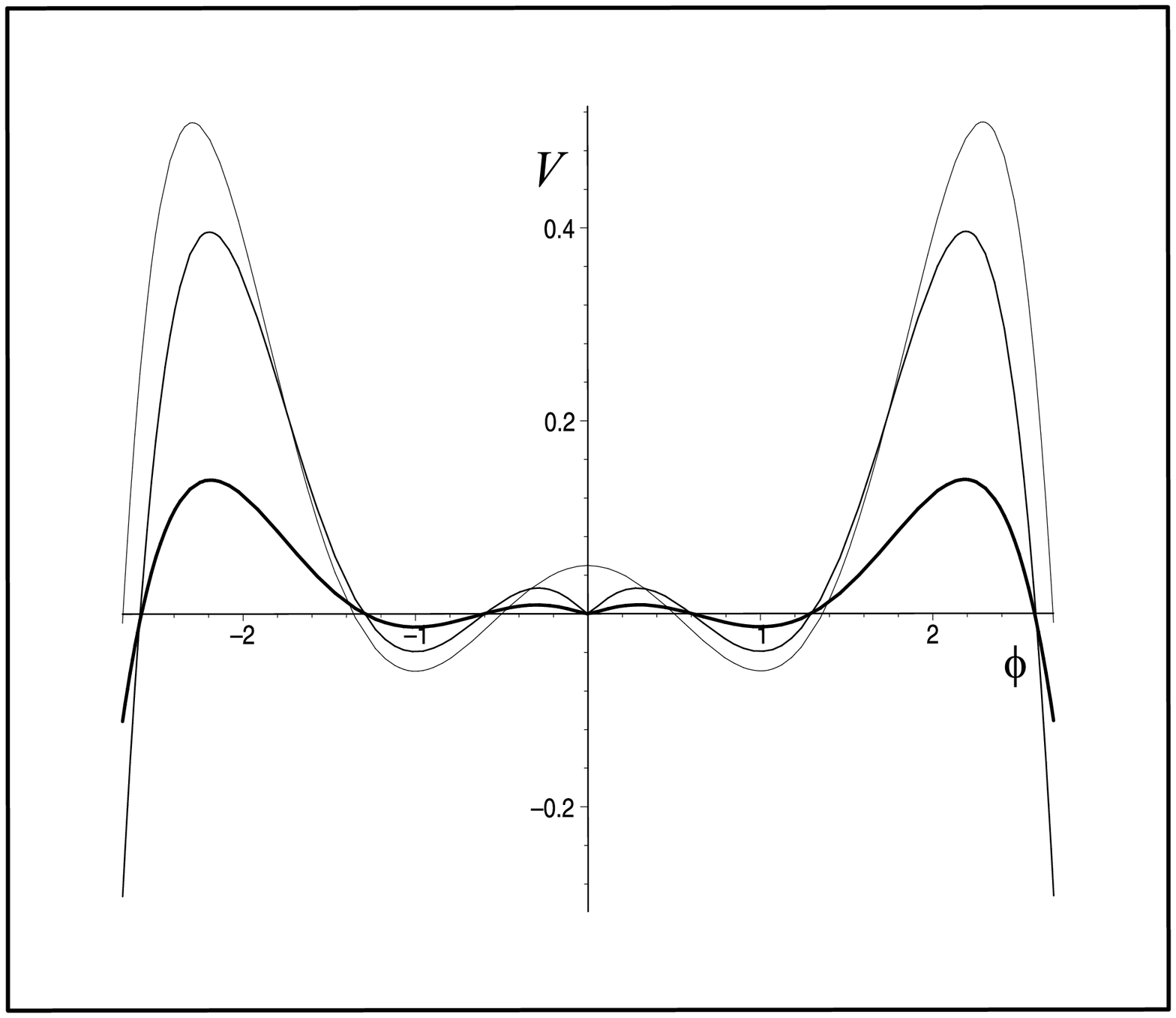,width=10.0cm}{Plots of the potential $V_p(\phi)$
for $p=1,3,\;{\rm and}\;5.$ Here and in the other figures the thickness
of the lines increases with increasing $p.$}
%%%%%%%%%%%%%%%%%%%%%%%%%%%%%%%%%%%%%%%%%%%%%%%%%%%%%%%%%%%%%%%%%%%%%%%

We return to curved spacetime, and we recall
that potentials defined as in Eq.~(\ref{gpot}), through the introduction
of the function $W=W(\phi)$, appear very naturally in supergravity,
and there $W$ is named superpotential --- see Refs.{\cite{bcy,cl,bbn}}
for more details on this subject.

The use of $W_p(\phi)$ to define the potential provides an important approach,
which leads to first-order equations that solve the equations of
motion --- see for instance Refs.~{\cite{cg,st,df}}. The first-order
equations are given by
\be\label{e1a}
\phi^{\prime}=\frac12\,\frac{dW_p}{d\phi}
\ee
and
\be\label{e1b}
A^\prime=-\frac13\,W_p
\ee
The first-order equation for $\phi$ reproduces the first-order equation
for the scalar field in flat spacetime \cite{bmm}, apart from the numerical
factor of $1/2$. We solve this equation to get \cite{bmm}
\be\label{phi}
\phi_p(y)=\tanh^p\left(\frac{y}{p}\right)
\ee
We see that for $p=1$ we get the standard solution. However, for
$p=3,5,...$ we get 2-kink solutions, which we show in Fig.~[2] --- see
Ref.~{\cite{bmm}} for more details on such defect structures.

%%%%%%%%%%%%%%%%%%%%%%%%%%%%%%%%%%%%%%%%%%%%%%%%%%%%%%%%%%%%%%%%%%%%%%%
\EPSFIGURE[!h]{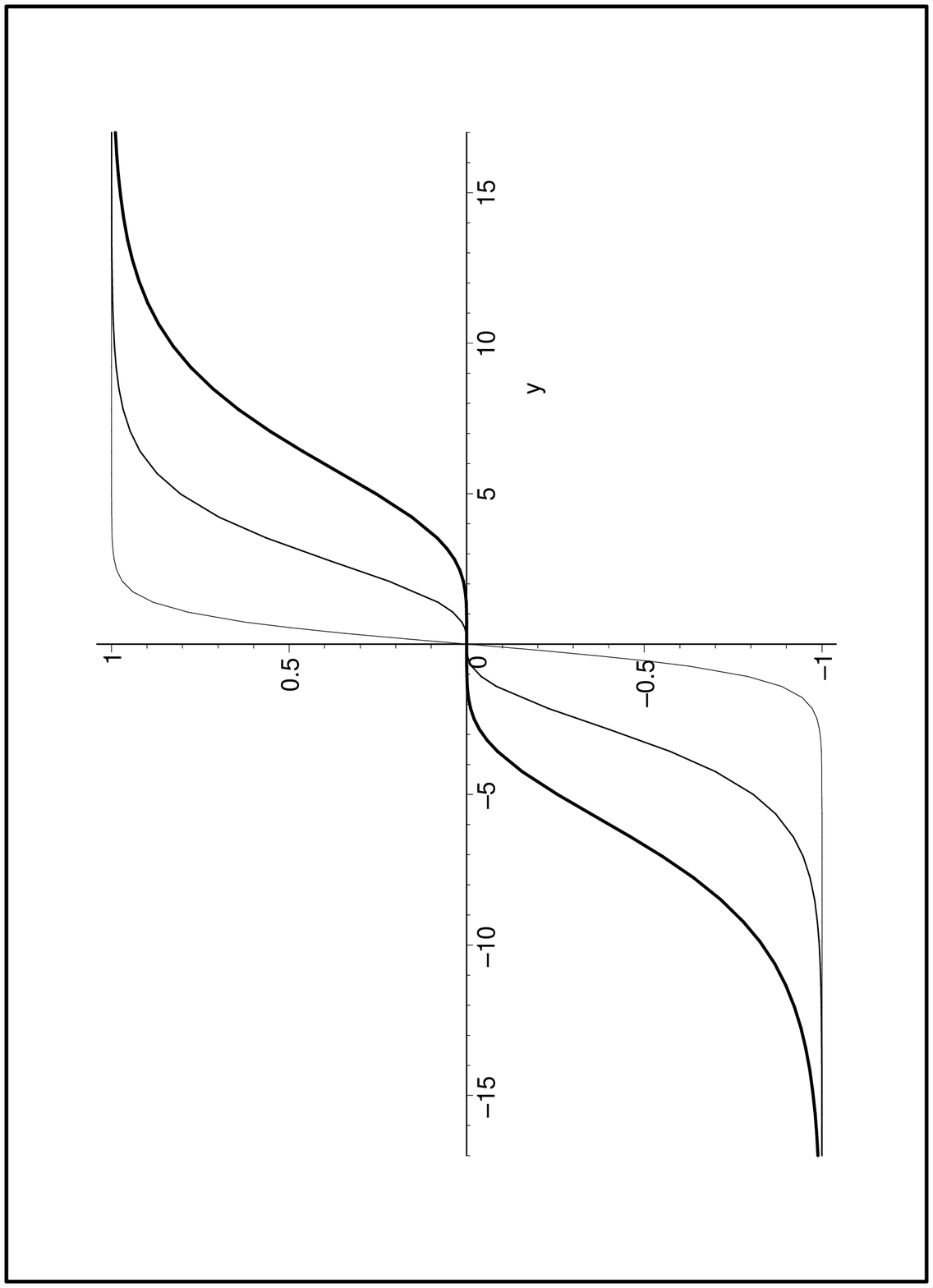,width=7.0cm,angle=270}{Plots of the solutions
$\phi_p(y)$ for $p=1,3,\;{\rm and}\;5.$}
%%%%%%%%%%%%%%%%%%%%%%%%%%%%%%%%%%%%%%%%%%%%%%%%%%%%%%%%%%%%%%%%%%%%%%%

The first-order equation (\ref{e1b}) for the warp factor is new for
$W_p$ given by Eq.~(\ref{w}), and we have been able
to solve it for $A_p(y)$ analytically. We get
\ben\label{warp}
A_p(y)\!\!&=&\!\!-\frac13\frac{p}{2p+1}\tanh^{2p}\left(\frac{y}{p}\right)-
\frac23\left(\frac{p^2}{2p-1}-\frac{p^2}{2p+1}\right)\nonumber
\\
&&\biggl{\{}\ln\biggl[\cosh\left(\frac{y}{p}\right)\biggr]-
\sum_{n=1}^{p-1}\frac1{2n}\tanh^{2n}\left(\frac{y}{p}\right)\biggr{\}}
\een
We plot the warp factor $e^{2A_p(y)}$ corresponding to these solutions
for $p=1,3,5$ in Fig.~[3].

The scalar matter field and the warp factor contribute to the brane scenario
according to the solutions (\ref{phi}) and (\ref{warp}). We use these results
to obtain the matter energy density $T_{00}^{\,p}(y)$ as a function of the
extra dimension. The analytic expression for $T_{00}^{\,p}$ is rather
awkward, thus we decided to plot it in Fig.~[4] to show the matter
behavior when $\phi(y)$ and $A(y)$ solve the first order equations
(\ref{e1a}) and (\ref{e1b}), respectively.

%%%%%%%%%%%%%%%%%%%%%%%%%%%%%%%%%%%%%%%%%%%%%%%%%%%%%%%%%%%%%%%%%%%%%%%
\EPSFIGURE[!h]{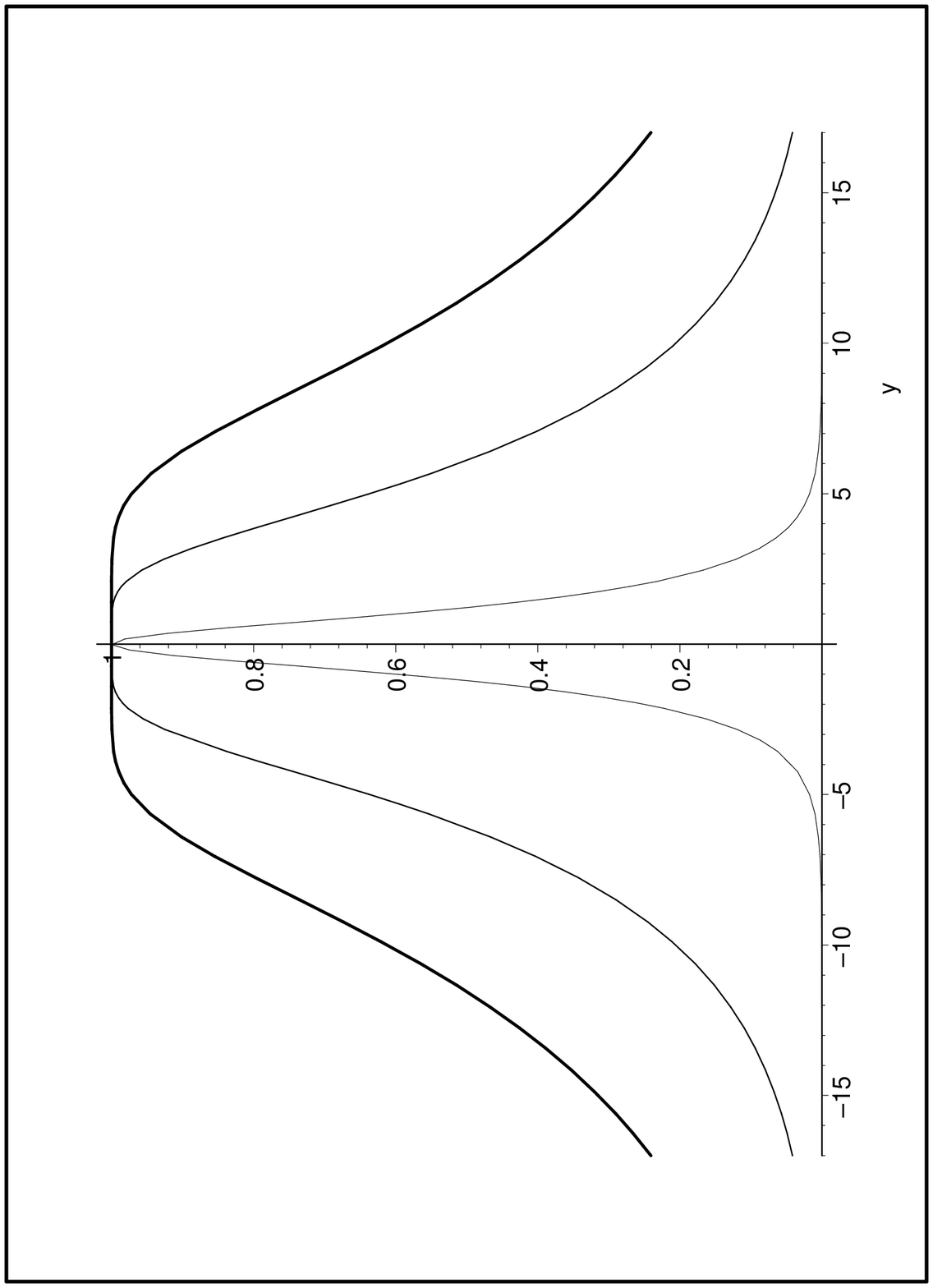,width=8.0cm,angle=270}{Plots of the warp factor
$e^{2 A_p(y)}$ for $p=1,3,\;{\rm and}\;5.$}
%%%%%%%%%%%%%%%%%%%%%%%%%%%%%%%%%%%%%%%%%%%%%%%%%%%%%%%%%%%%%%%%%%%%%%%

The plots depicted in Fig.~[4] show that the matter field gives
rise to thick branes composed of a single $(p=1)$ or two $(p=3,5,...)$
interfaces. In the last case, for $p=3,5,...,$ one sees the appearance
of a new phase in between the two interfaces where the energy density
of the matter field gets more concentrated. The scenario for $p=3,5,...,$
is that of a brane that supports internal structure \cite{m,brs,gp}.
We use Figs.~[3] and [4] to see that the warp factor is essencially constant
(for $p=3,5,...)$ in the very inside of the brane, in the region where
the energy density vanishes. This unveils the presence of a new phase
inside the brane, protected by the two interfaces that appear centered
at the points which maximize the energy density. For $p=1$ the scenario
is quite different, and the brane does not support the internal structure
that we clearly see for $p=3,5,...$

%%%%%%%%%%%%%%%%%%%%%%%%%%%%%%%%%%%%%%%%%%%%%%%%%%%%%%%%%%%%%%%%%%%%%%%
\EPSFIGURE[!h]{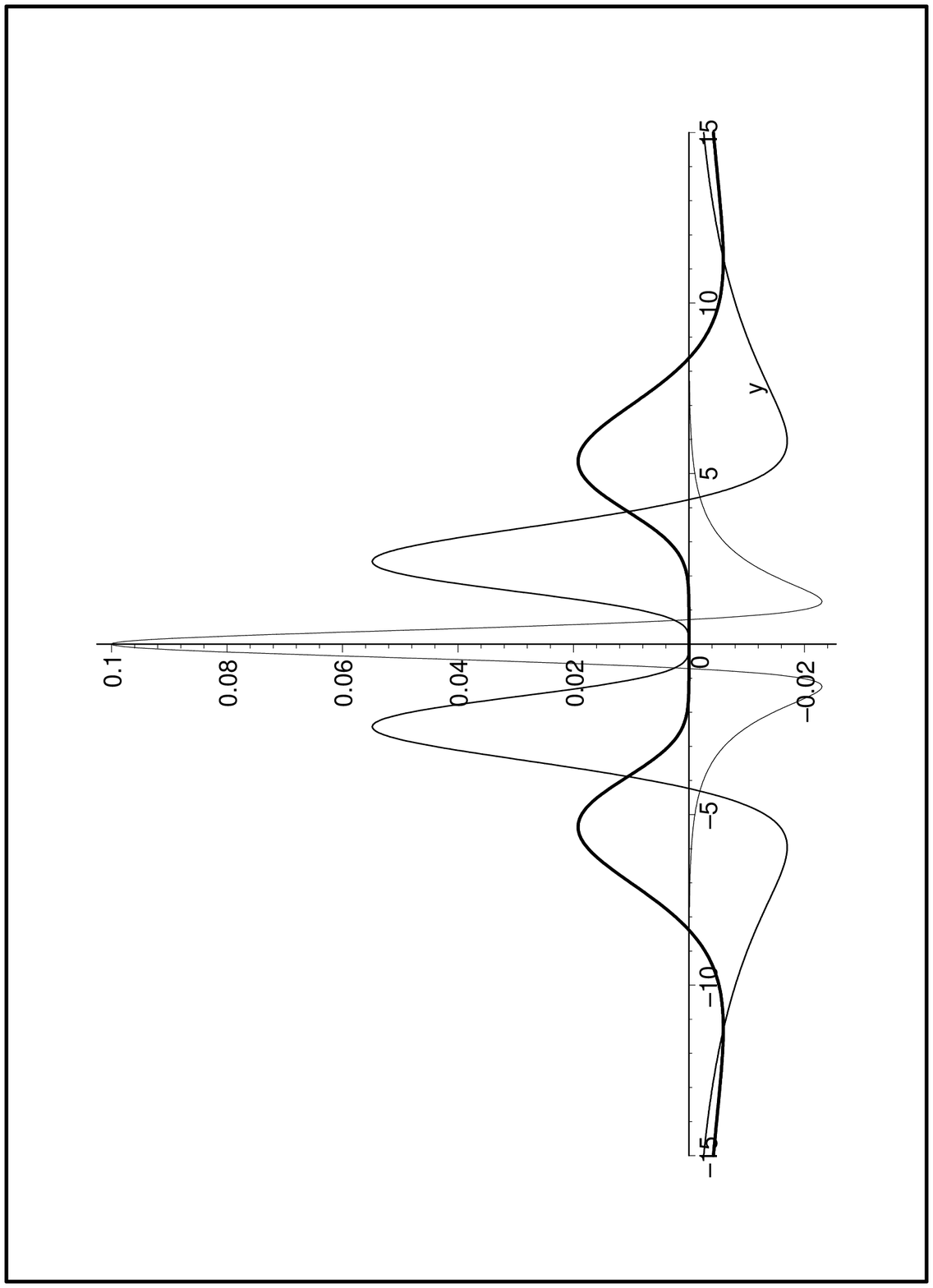,angle=270,width=9.0cm}{Plots of the matter energy
density $T^{\,1}_{00}(y)/10,$ $T^{\,3}_{00}(y),$ and
$T^{\,5}_{00}(y)$ for $p=1,3,\;{\rm and}\;5.$
We have plotted $1/10$ of $T^{\,1}_{00}$ to ease comparison.}
%%%%%%%%%%%%%%%%%%%%%%%%%%%%%%%%%%%%%%%%%%%%%%%%%%%%%%%%%%%%%%%%%%%%%%%

The scalar matter model (\ref{w}) that we have used leads to the solutions
(\ref{phi}) and (\ref{warp}), which are depicted in Figs.~[2] and [3],
respectively. We go further into the subject and we examine stability
of the above braneworld scenario. The key issue here is stability of
the geometry induced by the above solutions. We do this perturbing
the metric, using
\be
ds^2=e^{2A(y)}(\eta_{\mu\nu}+h_{\mu\nu})dx^\mu dx^\nu-dy^2
\ee
and supposing that $h_{\mu\nu}=h_{\mu\nu}(x,y)$ represent small perturbations.
Here we follow \cite{df} and we choose $h_{\mu\nu}$ in the form of transverse
and traceless contributions, ${\bar h}_{\mu\nu}.$ In this case one gets
the equation 
\be\label{h}
{\bar h}_{\mu\nu}^{\prime\prime}+4\,A^{\prime}
\,{\bar h}_{\mu\nu}^{\prime}=e^{-2A}\,\Box\,{\bar h}_{\mu\nu}
\ee
which describes linearized gravity. Here $\Box$ stands for the wave
operator in $(3,1)$ dimensions. We notice that in this sector,
gravity decouples from the matter field.    

We use the new coordinate $z$, which is defined by $dz=e^{-A(y)}dy$. Also,
we set
\be
{\bar h}_{\mu\nu}(x,z)=e^{ik\cdot x}e^{-\frac{3}{2}A(z)}H_{\mu\nu}(z)
\ee
In this case the above equation (\ref{h}) becomes the Schr\"odinger-like
equation
\be\label{se}
-\frac{d^2H_{\mu\nu}}{dz^2}+U_p(z)\,H_{\mu\nu}=k^2\,H_{\mu\nu}
\ee
where the potential is given by
\be
U_p(z)=\frac32\,A_p^{\prime\prime}(z)+\frac94\,A_p^{\prime2}(z)
\ee
The Hamiltonian in eq.~(\ref{se}) can be written in the form
\be
H=\left(-\frac{d}{dz}-\frac32A^{\prime}_p\right)
\left(\frac{d}{dz}-\frac32A^{\prime}_p\right)
\ee
and this ensures that $k$ is real, and $k^2\geq0;$ thus, there is no
unstable tachyonic excitation in the system. We solve for the zero
modes $(k=0)$ to get
\be
H^{(p)}_{\mu\nu}(z)=N^{(p)}_{\mu\nu}e^{3A_p(z)/2}
\ee
where $N^{(p)}_{\mu\nu}$ is a normalization factor. In Fig.~[5]
we plot the potentials $U_1(z)/10, U_3(z),$ and $U_5(z)$, and the
corresponding zero modes for $p=1,3,5.$ In Fig.~[6] we plot wave functions
for massive modes in the two regions where $k^2<{\rm max}[U_p(z)]$ and
$k^2>{\rm max}[U_p(z)].$ These wave functions confirm the expected result,
that they describe motion in the bulk, not bound to the brane.

%%%%%%%%%%%%%%%%%%%%%%%%%%%%%%%%%%%%%%%%%%%%%%%%%%%%%%%%%%%%%%%%%%%%%%%
\begin{figure}[ht!]
\includegraphics[{angle=270,width=7.5cm}]{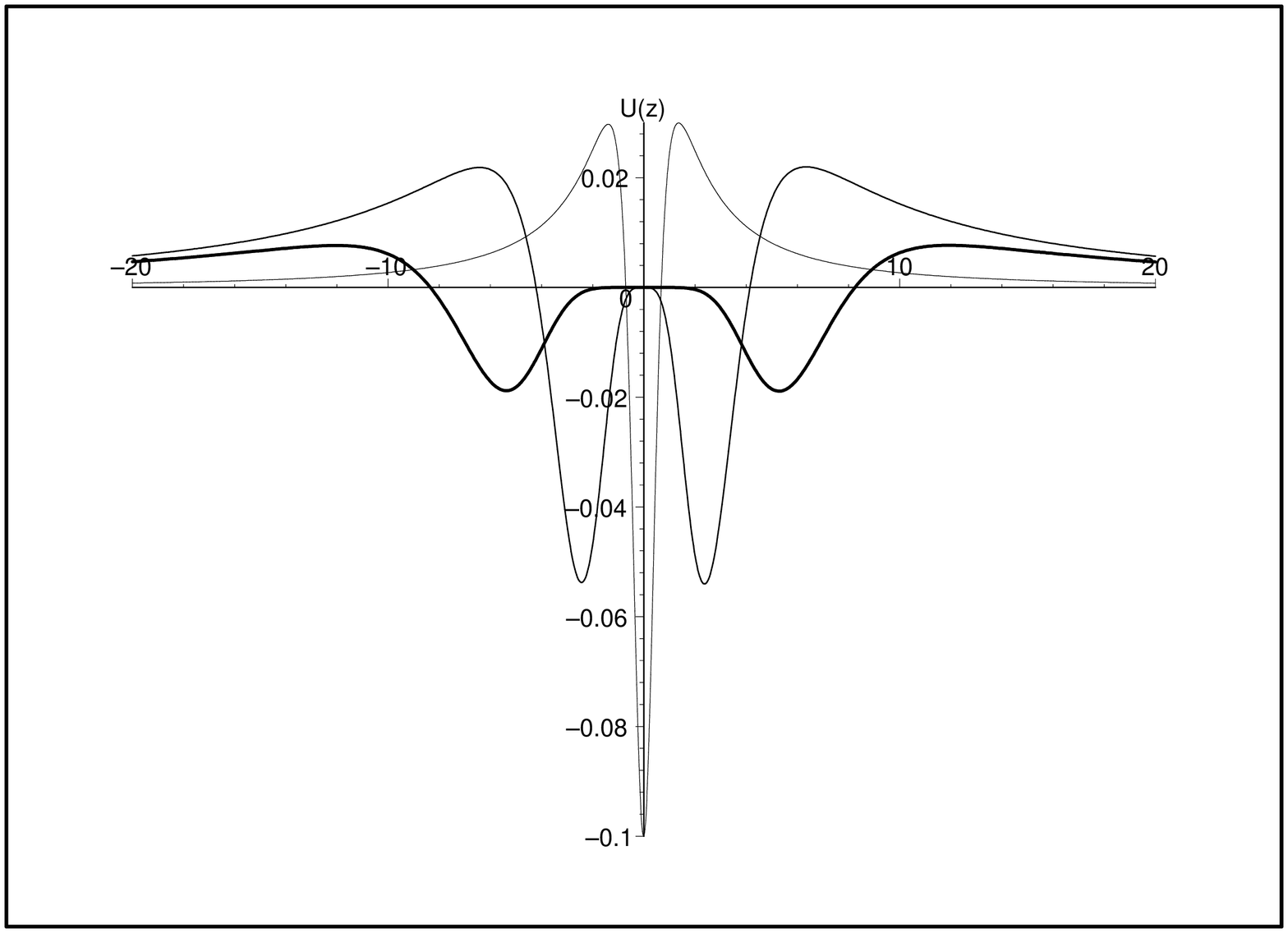}
\includegraphics[{angle=270,width=7.5cm}]{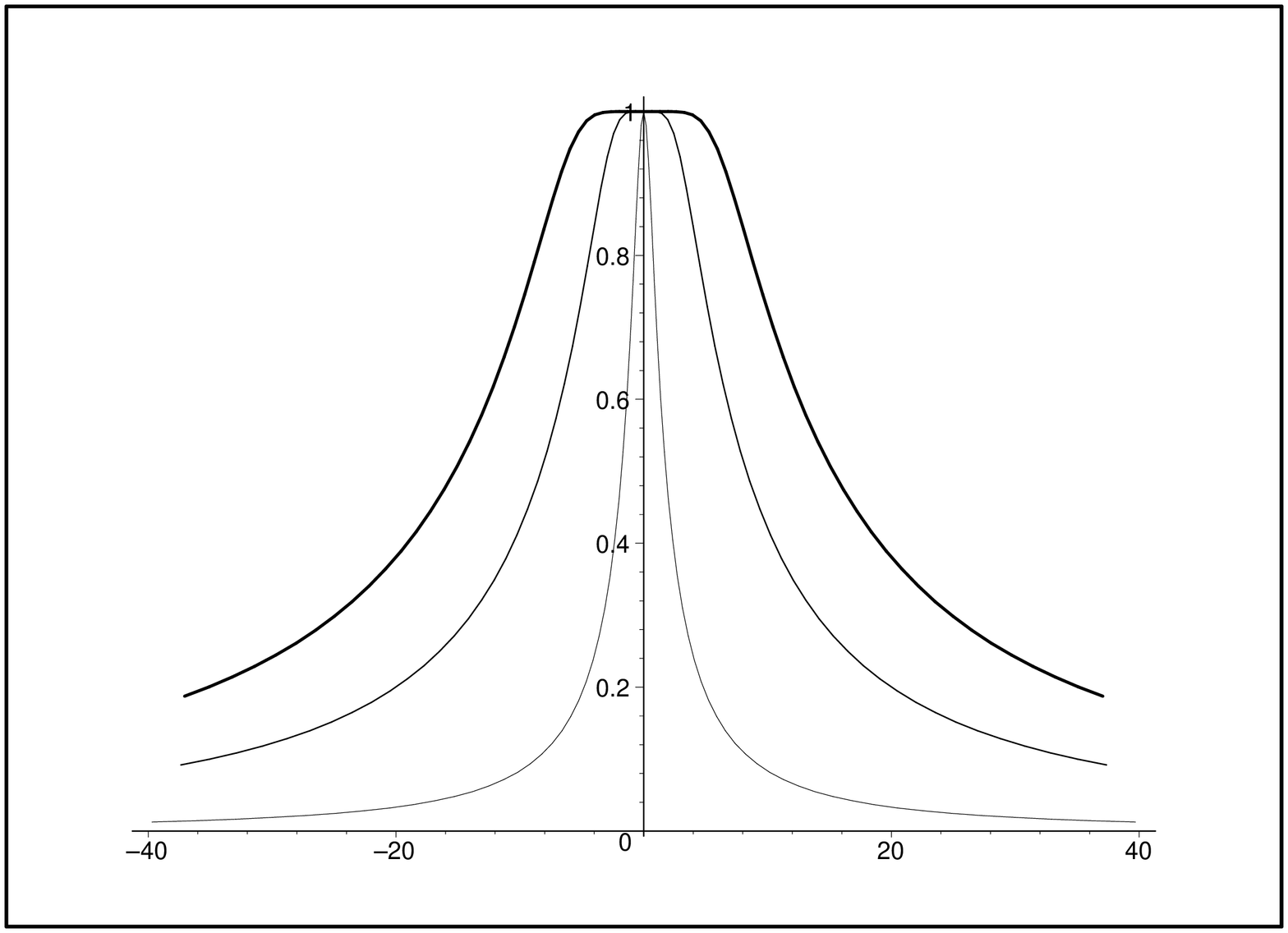}
\caption{Plots of the potentials $U_1(z)/10, U_3(z),\, {\rm and}\, U_5(z)$
(left) and the corresponding zero modes (right). We have plotted $U_1(z)/10$
to ease comparison.}
\end{figure}
%%%%%%%%%%%%%%%%%%%%%%%%%%%%%%%%%%%%%%%%%%%%%%%%%%%%%%%%%%%%%%%%%%%%%%%
%%%%%%%%%%%%%%%%%%%%%%%%%%%%%%%%%%%%%%%%%%%%%%%%%%%
\begin{figure}[!h]
\includegraphics[{angle=270,width=7.5cm}]{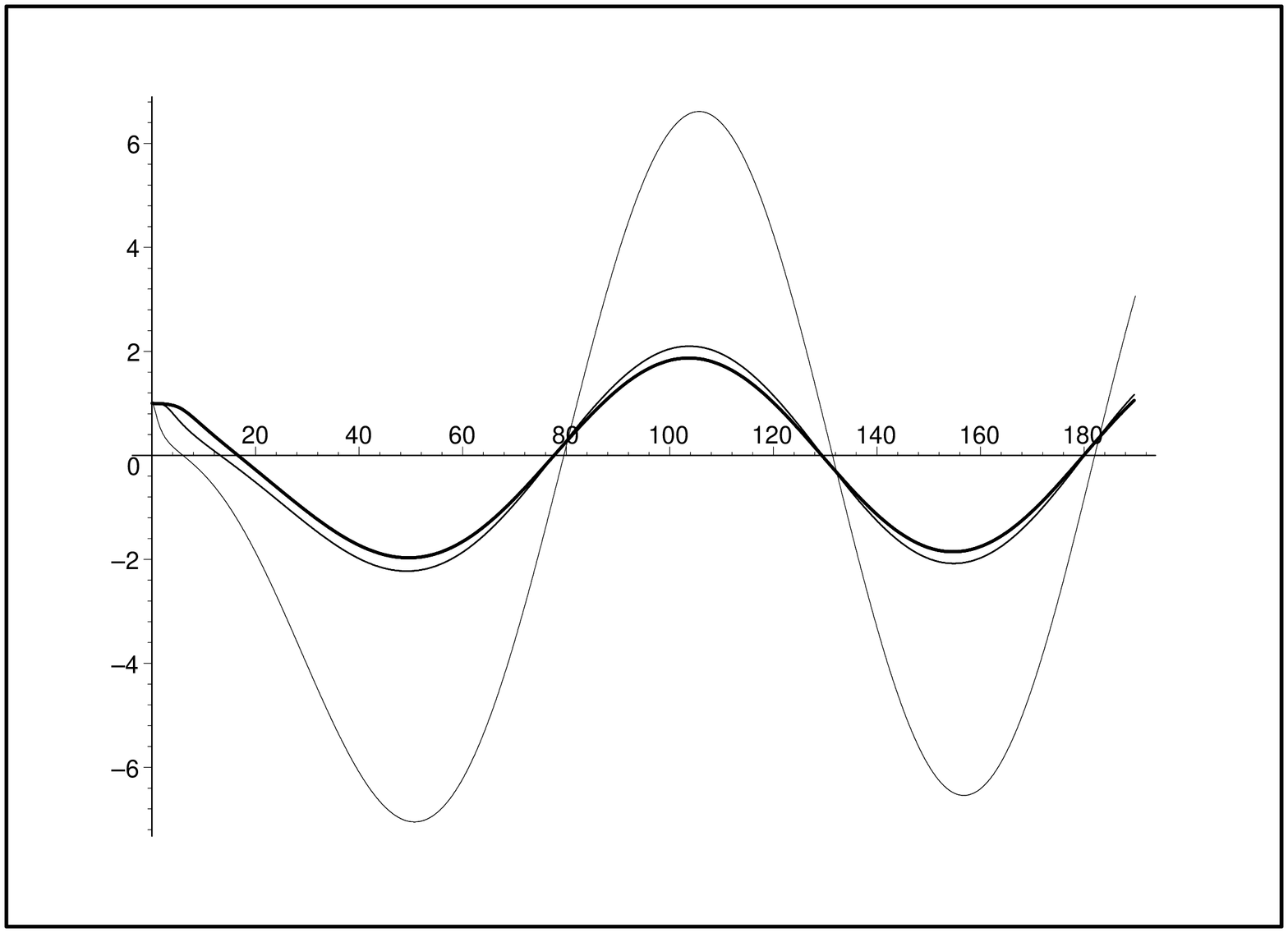}
\includegraphics[{angle=270,width=7.5cm}]{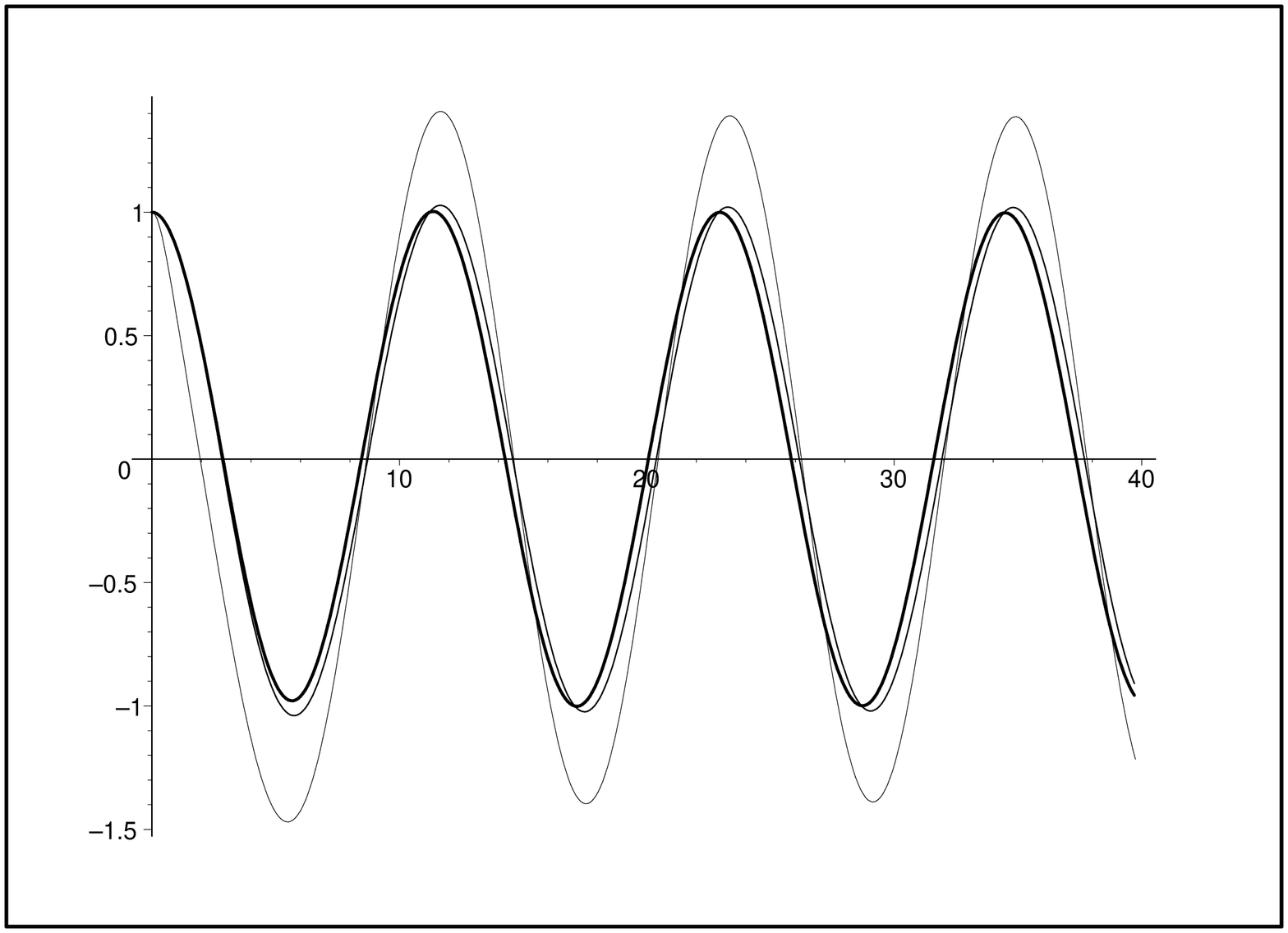}
\caption{Plots of wave functions for nonzero modes for 
$k^2<{\rm max}[U_p(z)]$ (left) and for $k^2>{\rm max}[U_p(z)]$ (right)
for $p=1,3,\;{\rm and}\;5.$}
\end{figure}
%%%%%%%%%%%%%%%%%%%%%%%%%%%%%%%%%%%%%%%%%%%%%%%%%%%%%%%%%%%%%%%%%%%%%%%

In the brane scenario that we have just examined, in order for
the zero modes to describe localized four dimensional gravity, normalizability
is essential. To ensure normalizability, the zero modes as a function of $z$
must fall off faster than $z^{-1/2}.$ For this reason, in Fig.~[7a] we have
plotted the zero modes and their intersection points with the function
$f(z)=z^{-1/2}.$ And in Fig.~[7b] we show that the intersection points
$z_p^i$ increases with $p$ with decreasing derivative. This result indicates
that the zero modes always fall off faster than $z^{-1/2}$, showing that they
are normalizable, giving rise to localized four dimensional gravity for
every $p$ odd integer.

A consequence of this result is that the massive modes are strongly suppressed.
The structure of the volcano potentials plotted in Fig.~[5a] could contribute
to the appearance of resonances for $k<{\rm max}[U_p(z)]$. To check for the
presence of resonances, we have investigated massive states for several
different values of $k$ in the interval $[0,{\rm max}[U_p(z)]],$ for $p=3$
and $5.$ We have found no evidence for resonances. Thus, our model seems
to support no resonance for $p$ odd integer, despite the fact that the values
$p=3,5,...$ lead to the generation of two interfaces which inhabit the thick
brane presented in this work.

%%%%%%%%%%%%%%%%%%%%%%%%%%%%%%%%%%%%%%%%%%%%%%%%%%%%%%%%%%%%%%%%%%%%%%%
\begin{figure}[ht!]
\includegraphics[{angle=270,width=7.5cm}]{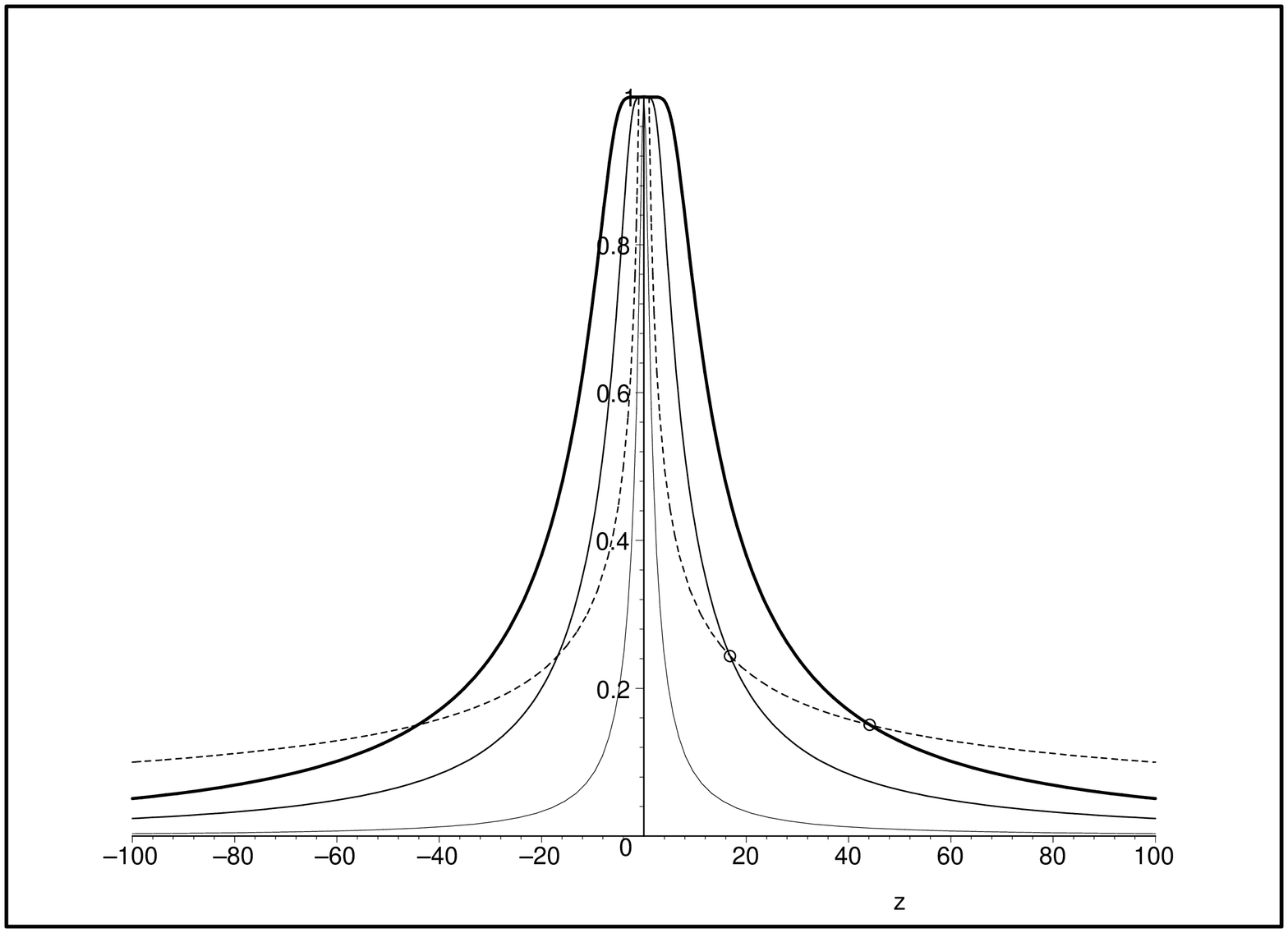}
\includegraphics[{angle=270,width=7.5cm}]{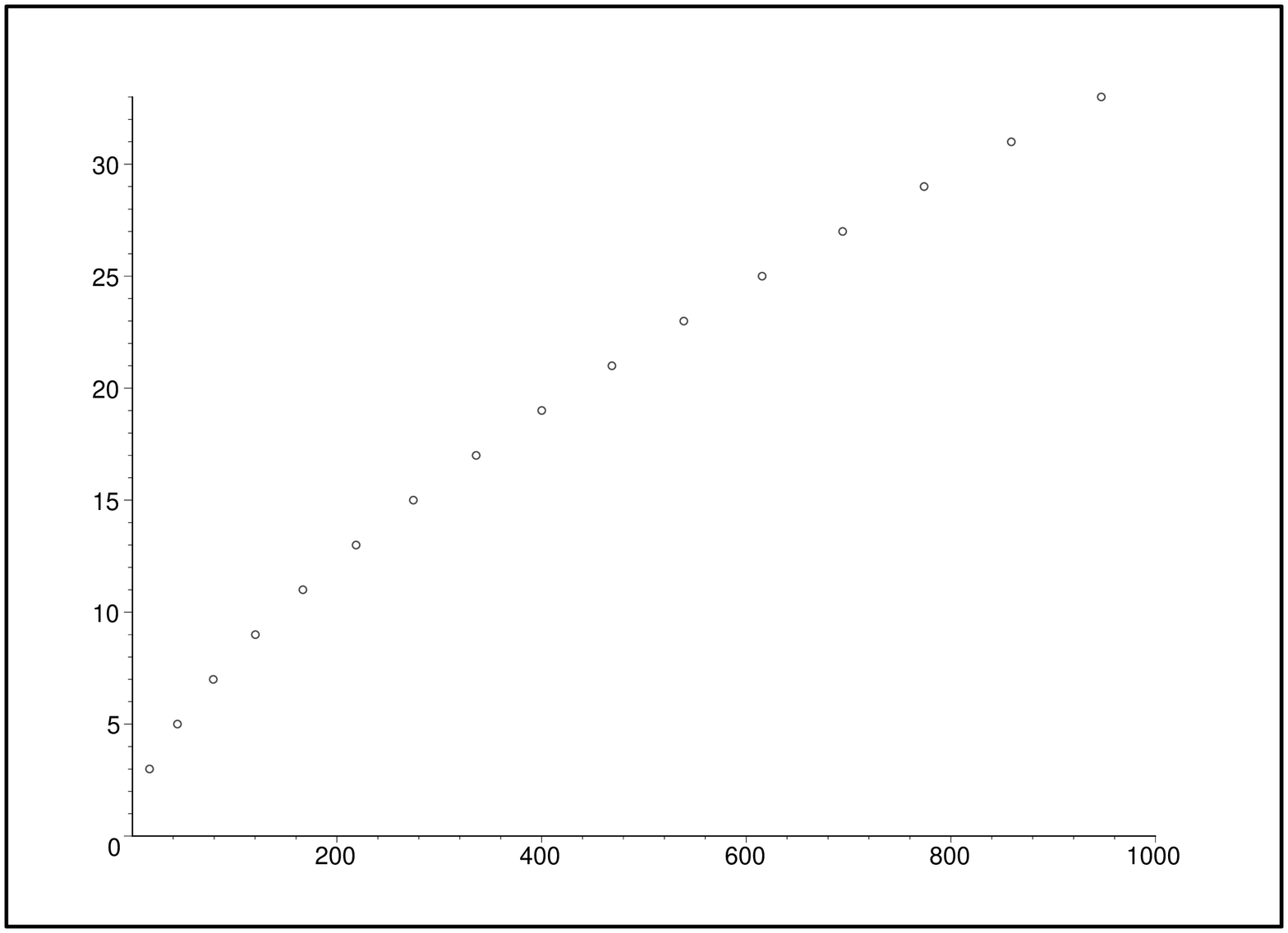}
\caption{Plots of the zero modes for $p=1,3,5$ (left),
and intersection points with $f(z)=z^{-1/2},$ which is shown 
as the dashed line. We also plot (right) the intersection points between
the zero modes and the function $f(z)=z^{-1/2}$ for several values of $p.$}
\end{figure}
%%%%%%%%%%%%%%%%%%%%%%%%%%%%%%%%%%%%%%%%%%%%%%%%%%%%%%%%%%%%%%%%%%%%%%%

Let us now comment a little further on the 2-kink solutions that we have
found in Eq.~(\ref{phi}). We can have a better view of these defect structures
calculating ${\wt y}_p$, the position which identifies the center
of the defect. We identify ${\wt y}_p$ with the maxima of the derivative
of the defect solutions, that is, ${\wt y}_p=\max[d\phi_p(y)/dy].$ We get
${\wt y}_p=\pm p\,{\rm arctanh}\,\sqrt{(p-1)/(p+1)}.$
For $p=1$ it gives ${\bar y}_1=0,$ as expected, but for $p=3,5,...,$ we obtain
two non-zero values, which are directly related to the center of the interfaces
that inhabits the defect. In flat spacetime, the values ${\wt y}_p$
exactly identify the maxima of the energy density
\be
{\wt T}^p_{00}(y)=\frac12\left(\frac{d\phi_p}{dy}\right)^2 +
\frac12 \left(\frac12\frac{d W_p}{d\phi}\right)^2
\ee
for the corresponding solutions $\phi_p(y)$ --- the unusual $1/2$ factor
that multiplies $W_p$ appears because in flat spacetime we have to redefine
$W_p$ as $W_p/2,$ in order not to change the first-order equation (\ref{e1a})
for the scalar field $\phi.$ This result shows that the 2-kink
solutions are composed of two interfaces, each one being centered at each one
of the two points ${\wt y}_p$ that we have just obtained.
In curved spacetime, however, the position of the maxima of the matter energy
density is different. To see this, we use $T^p_{00}(y)$ to identify the maxima
of the matter energy density in curved spacetime. The function $T^p_{00}(y)$
is depicted in Fig.~[4] for some values of $p.$ There are maxima at the points
$y_3=2.42,\,y_5=5.36,\,y_7=8.70,...,$ which are different from the points
${\wt y}_3=2.64,\,{\wt y}_5=5.73,\,{\wt y}_7=9.22,...,$
which identify the maxima of the energy density in flat spacetime, or
the center of the interfaces of the 2-kink solutions that we have
already found. This result leads to the conclusion that the curved spacetime
induces an attraction between the interfaces that appear from the 2-kink
solutions. We have investigated this effect for several $p$, and we could
verify that the relative attraction decreases for increasing $p.$

We recall that solutions similar to the 2-kink defects
depicted in Fig.~[2] have already been considered before in flat spacetime.
See for instance Ref.~{\cite{chw}}, which investigates a model described
by a complex scalar field. We can also deal with two real scalar fields
and consider the model investigated in \cite{bds} in flat spacetime.
It is defined by $W(\phi,\chi)=\phi-\phi^3/3-r\phi\chi^2,$ where $r$
is a real parameter. It supports kink-like solutions \cite{sv,ilg}
of the form presented in Fig.~[2]. This model was very recently used
to describe matter coupled to gravity in warped geometry \cite{es},
with the result that the warp factor resembles the form depicted in
Fig.~[3]. The models investigated in the present work are similar but
simpler, since they are described by a single real scalar field.

The model investigated in \cite{bds} support two-field
solutions in the sector defined by the minima $(\pm1,0)$ in the
$(\phi,\chi)$ plane \cite{bb}. We may couple this model to gravity,
to investigate whether these two-field solutions generate braneworld scenarios.
Other extensions may involve scalar fields described under the general
guidance of the $Z_2\times Z_3$ symmetry --- see for instance the second
work in Ref.~{\cite{bb}} --- and they may also lead to issues involving branes
that host network of non trivial structure.

We would like to thank F.A. Brito and J.R. Nascimento for helpful discussions.
We also thank PROCAD and PRONEX for financial support. DB and CF thank CNPq
for partial support and ARG thanks FAPEMA for a fellowship.

\end{document}